\documentclass[preprint,authoryear,12pt]{elsarticle}

\usepackage{amssymb}

\journal{Advances in Space Research}

\begin{document}

\begin{frontmatter}



\title{Probing the Nature of Short Swift Bursts via Deep INTEGRAL Monitoring of GRB 050925}


\author[cresstnasa,umbc]{T. Sakamoto}
\author[nasa]{L. Barbier}
\author[nasa]{S. D. Barthelmy}
\author[cresstnasa,umbc]{J. R. Cummings}
\author[lanl]{E. E. Fenimore}
\author[nasa]{N. Gehrels}
\author[cresstnasa,usra]{H. A. Krimm}
\author[cresstnasa,umcp]{C. B. Markwardt}
\author[lanl]{D. M. Palmer}
\author[nasa]{A. M. Parsons}
\author[isas]{G. Sato}
\author[orau]{M. Stamatikos}
\author[nasa]{J. Tueller}

\address[cresstnasa]{CRESST and NASA Goddard Space Flight Center, Greenbelt, MD 20771}
\address[umbc]{Joint Center for Astrophysics, University of Maryand, Baltimore County, Baltimore, MD 21250}
\address[nasa]{NASA Goddard Space Flight Center, Greenbelt, MD 20771}
\address[lanl]{Los Alamos National Laboratory, P.O. Box 1663, Los Alamos, NM, 87545}
\address[usra]{Universities Space Research Association, Columbia, MD 21044}
\address[umcp]{Department of Astronomy, University of Maryland, College Park, MD 20742}
\address[isas]{Institute of Space and Astronautical Science, JAXA, Kanagawa 229-8510, Japan}
\address[orau]{Oak Ridge Associated Universities, P.O. Box 117, Oak Ridge, Tennessee 37831}

\begin{abstract}
We present results from {\it Swift}, {\it XMM-Newton}, and deep INTEGRAL monitoring 
in the region of GRB 050925.  This short {\it Swift} burst is a candidate for a newly discovered soft gamma-ray 
repeater (SGR) with the following observational burst properties: 1) galactic plane (b=$-0.1^{\circ}$) 
localization, 2) 150 msec duration, and 3) a blackbody rather than a simple power-law spectral 
shape (with a significance level of 97\%).  We found two possible X-ray counterparts of GRB 050925 
by comparing the X-ray images from {\it Swift} XRT and {\it XMM-Newton}.  Both X-ray sources show 
the transient behavior with a power-law decay index shallower than $-1$.  We found no hard X-ray 
emission nor any additional burst from the location of GRB 050925 in $\sim$5 Ms of {\it INTEGRAL} 
data.  We discuss about the three BATSE short bursts which might be associated 
with GRB 050925, based on their location and the duration.  Assuming GRB 050925 
is associated with the H$_{II}$ regions (W 58) at the galactic longitude 
of $l$=70$^{\circ}$, we also discuss the source frame properties of GRB 050925.  

\end{abstract}


\end{frontmatter}


\section{Introduction}

The origin of the short ($<$ 2 seconds) class of gamma-ray bursts (GRBs) 
is receiving huge attention in the field of high-energy astrophysics.  
Thanks to the rapid position notices and response by {\it HETE-2} and $Swift$, 
afterglow emission has been found for a handful of short GRBs \citep[e.g.,][]{berger2007}.  
Less than arc-second positions, which are provided by X-ray and optical afterglows, enable 
us to study the host galaxies of short GRBs.  Surprisingly, unlike the 
long duration GRBs which always have host galaxies with a high star forming rate \citep{bloom2002}, 
short GRBs emerge from both star-forming and non-star forming 
galaxies \citep[e.g.,][]{villasenor2005}.  This suggests that a substantial range of lifetimes  
is needed for the progenitors of short GRBs.  This discovery tightens the case 
for a different origin for short and long GRBs.  

On the other hand, some fraction of short bursts might be from local 
and extra galactic flares of soft gamma-ray repeaters (SGRs).  SGRs 
are believed to be highly magnetized isolated neutron stars \citep{dt1992}.  They produce 
short spikes (a few tens of milliseconds) \citep{woods2002} and sometimes a giant flare 
\citep{hurley1999,palmer2005} in $\gamma$-rays.  Although it is possible to 
detect giant flares of extragalatic SGRs from nearby galaxies, their fraction of 
such flares among short GRBs is still not clear \citep[e.g.,][]{palmer2005,lazzati2005}, 
Furthermore, a small flare from a previously unknown SGR in the galaxy 
might be detected as a short burst in $\gamma$-rays.  

\section{GRB 050925}
\subsection{Swift/BAT Prompt Emission}

\begin{figure}
\begin{center}
\includegraphics[height=12cm,clip,angle=-90]{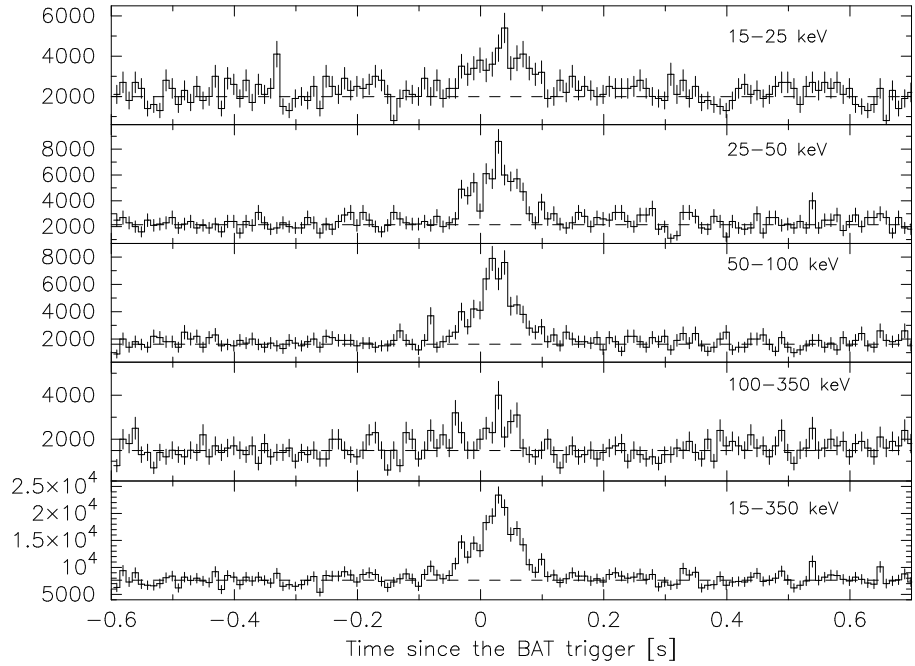}
\caption{\label{bat_lc} The BAT five channel light curves in 
10 msec binning.  The background has not been subtracted.}
\end{center}
\end{figure}

On 25 September 2005, the ${\it Swift}$ Bust Alert Telescope (BAT) 
instrument detected GRB 050925, which only lasted for $\sim$100 ms and 
possibly consists of two pulses (Figure \ref{bat_lc}).  The BAT ground 
analysis position of this burst is 
(R.A., Dec.) = (20$^{\rm h}$ 13$^{\rm m}$ 56.9$^{\rm s}$, 
34$^{\circ}$ 19$^{\prime}$ 48$^{\prime\prime}$) (J2000) which corresponds to 
the galactic coordinate of ($l$, $b$) = ($72.320^{\circ}$, $-0.101^{\circ}$) \citep{sakamoto2010}.  
Its location is in the galactic plane ($b= -0.1^{\circ}$).  The $T_{90}$ and 
$T_{50}$ duration is 90 ms and 40 ms respectively.  
The fluence in the 104 ms time window ($T_{100}$ interval) in the 15-150 keV band is 
$(7.7 \pm 0.9)$ $\times$ 10$^{-8} $ ergs cm$^{-2}$.  The peak flux in the 10 ms time 
window is estimated to be $(1.2 \pm 0.1) \times 10^{-6}$ ergs cm$^{-2}$ s$^{-1}$.  

\begin{figure}
\begin{center}
\includegraphics[height=6cm,clip]{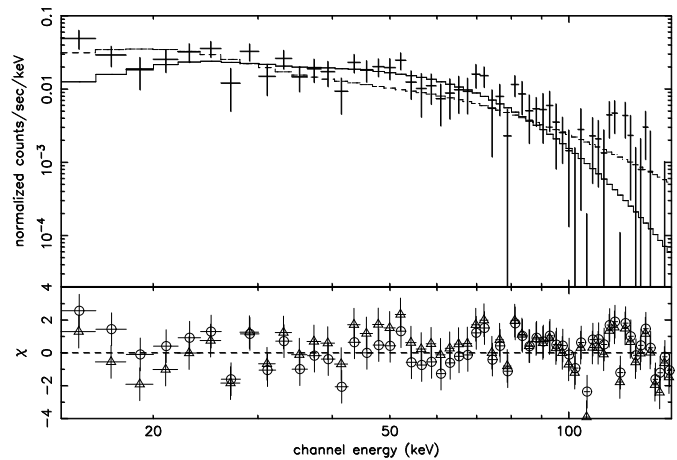}
\caption[short caption for figure3]{\label{bat_spec} The BAT time-integrated spectrum of the 
burst (104 msec duration).  The best fit with a blackbody model is shown in a solid line (upper) 
with circles in a residual panel (bottom), whereas, the best fit with a simple 
power-law model is shown in a dashed line (upper) with triangles in a residual panel (bottom). }
\end{center}
\end{figure}

\begin{figure}
\begin{center}
\includegraphics[height=9.5cm,clip,angle=-90]{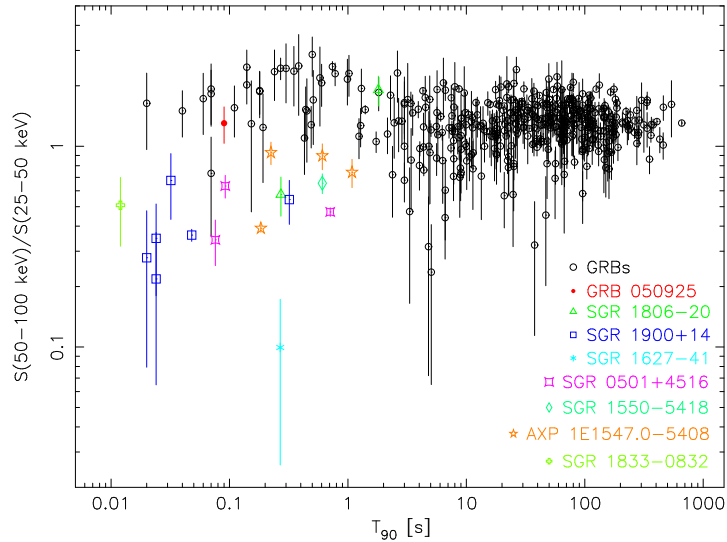}
\caption[short caption for figure2]{\label{dur_hr} The BAT fluence ratio between the 50-100 keV and 
25-50 keV band vs. $T_{90}$ duration.  The BAT GRBs (black open circles), GRB 050925 (red filled circle), 
and short bursts from SGRs/AXPs, which are detected by BAT, are shown in different marks and colors.  
}
\end{center}
\end{figure}

As seen in Figure \ref{bat_spec}, the prompt emission spectrum of this burst 
shows a better fit to a blackbody (BB) spectrum with a temperature of 15.1 $\pm$ 0.5 keV 
($\chi^{2}$/dof = 75.3/57) over a simple power-law (PL) spectrum ($\chi^{2}$/dof = 87.7/57).  
To quantify the significance of this improvement, we performed 10,000 spectral 
simulations assuming the best-fit spectral parameters in a simple power-law 
model and determined in how many cases the blackbody fit gives $\chi^{2}$ improvements 
of equal or greater than $\Delta$$\chi^{2}$ = $\chi^{2}$(PL) - $\chi^{2}$(BB) = 12.4 
over the simple power-law model. We found equal or higher improvements in $\chi^{2}$ 
in 691 simulated spectral out of 10,000.  Thus, the chance probability of having an 
equal or higher $\Delta$$\chi^{2}$ of 12.4 with the blackbody model when the parent 
distribution is a simple power-law model is 7\%.  A blackbody with a 
temperature of $\sim$ 10 keV 
is a typical spectral model for short SGR bursts in 
the BAT energy range \citep[e.g.,][]{olive2005,fenimore1994}.  
GRB 050925 belongs to a softer region compared to the other short GRBs detected by BAT 
in the hardness-duration plot (Figure \ref{dur_hr}).  

The spectral lag using the 10 ms light curves between the 50-100 keV and the 
15-50 keV band is $0.012 \pm 0.055$ s (1 $\sigma$) which is consistent with zero.  
This negligible spectral lag of GRB 050925 is consistent with either 
short GRBs \citep[e.g.,][]{norris2010} or short bursts from SGRs \citep[e.g.,][]{nakagawa2007}.  

\subsection{X-ray Observations}

\begin{figure}[t]
\begin{center}
\includegraphics[height=9.0cm]{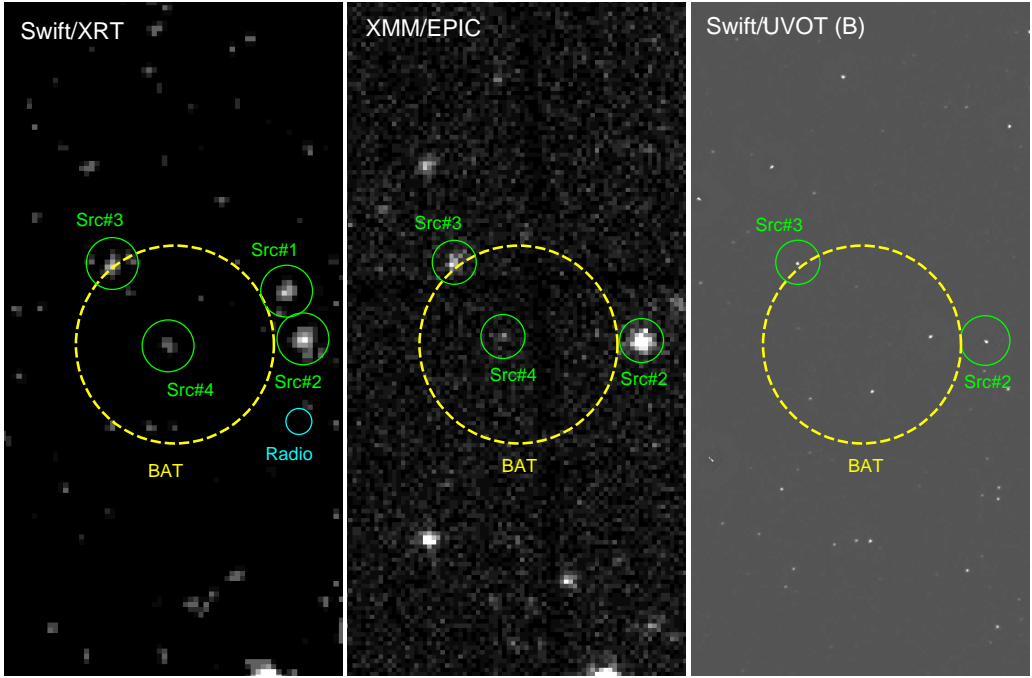}
\caption[short caption for figure r]{\label{xrt_xmm_uvot} The X-ray and B band images of 
the field of GRB 050925 (left: {\it Swift} XRT, middle: {\it XMM-Newton}, and right: {\it Swift} 
UVOT B filter).  The larger dashed yellow circle is the BAT 90\% error circle based on 
the ground analysis.  The location of a radio source, which mentioned in section 2.3, 
is shown in a cyan circle.}
\end{center}
\end{figure}

The ${\it Swift}$ X-ray Telescope (XRT) started observing the BAT error 
circle about 100 seconds after the burst with a net exposure of 30 ksec.  
XRT found four sources (labeled as Src \#1 to \#4 as shown in left panel of 
Figure \ref{xrt_xmm_uvot}) near and inside the BAT 90\% error circle.  
Only Src \#4 is located completely inside the BAT error circle.  The positions of 
Src \#2 and Src \#3 are coincident with stars in the guide star catalog (GSC) 
version 2.2 (N033223385350 and N033223389), and are also detected by ${\it Swift}$ 
UV-optical telescope (right panel of Figure \ref{xrt_xmm_uvot}).  Therefore, 
these two X-ray sources are stars, and not associated with GRB 050925.  

The {\it XMM-Newton} observed the field of GRB 050925, starting from 17 days after the 
BAT trigger, for 26.6 ksec \citep{rea2005}.  The XMM EPIC mosaic image (middle panel of Figure 
\ref{xrt_xmm_uvot}) shows detections of Src \#2, Src \#3 and Src \#4.  However, 
there is no detection of Src \#1 in the XMM image.  Based on the 
count-rate in the sensitivity map at the location of Src \#1, the upper-limit 
of the flux is $1 \times 10^{-14}$ erg cm$^{-2}$ s$^{-1}$ (0.2-12 keV).  Using 
the WebPIMMS\footnote{http://heasarc.gsfc.nasa.gov/Tools/w3pimms.html} and 
assuming the same absorbed power-law spectrum in the flux calculation 
of the XMM count-rate ($N_{H} = 3 \times 10^{20}$ and a power-law photon index of 1.7), 
the XRT count-rate of $(2.3 \pm 0.3) \times 10^{-3}$ c/s corresponds to $(1.05 \pm 0.14) 
\times 10^{-13}$ erg cm$^{-2}$ s$^{-1}$.  By combining the XRT detection 
and the corresponding upper-limit in the 
XMM observation of Src \#1, we found that the source has to decline steeper than 
a power-law index of $-0.7$.  Although Src \#1 is just outside the BAT error circle, 
its transient nature makes it a possible X-ray counterpart of GRB 050925.  

The XRT count-rate of Src \#4, which is the only X-ray source inside the BAT error circle, 
is $(7.0 \pm 1.7) \times 10^{-4}$ c/s.  The estimated absorbed flux using the same 
spectral parameters of Src \#1 is $(3.2 \pm 0.8) \times 10^{-14}$ erg cm$^{-2}$ s$^{-1}$.  
On the other hand, the estimated flux in the XMM EPIC mosaic image is $(4.2 \pm 2.4) 
\times 10^{-15}$ erg cm$^{-2}$ s$^{-1}$.  Therefore, the source shows a decline in 
the flux by a power-law index of $-0.6$ between these two observations.  

Based on analysis of these observations from XRT and XMM, both Src \#1 and Src \#4 show 
a decline in the flux between these two observations.  Therefore, at this stage, it is not 
possible to identify which source is the X-ray counterpart of GRB 050925.  

\subsection{Optical, Infrared and Radio Observations}

The 2-m Faulkes North Telescope followed up GRB 050925 at 3.3 minutes 
after the BAT trigger time.  No new optical source found inside the BAT flight 
location with the upper limit of 19 mag in R \citep{guidorzi2005}.  The PAIRITEL 
1.3-m telescope observed GRB 050925 at 17.6 hours after the trigger in infrared 
bands.  No new source was found within 1$^{\prime\prime}$ of the radio source 
(see below about the radio source) with 5 $\sigma$ upper limits of 18.3 mag, 18.5 mag 
and 19.3 mag in K$_{\rm S}$, J and H, respectively \citep{bloom2005}.  
{\it Swift} UVOT also found no new source 
within the BAT error circle with the 3 $\sigma$ upper limits of 20.2 mag, 21.8 mag and 
21.2 mag in V, B and U, respectively \citep{rosen2005}.

The Westerbork Synthesis Radio Telescope observed GRB 050925 in 4.2-16.3 hours 
after the BAT trigger and found no radio source within the BAT error circle with a 
3 $\sigma$ upper limit of 72 $\mu$Jy.  However, they note a bright radio source, 
which turns out to be a constant source based on 2nd observation ($\sim$6 days 
after the trigger), just outside the BAT error circle at (R.A., Dec.) = 
(20$^{\rm h}$ 13$^{\rm m}$ 47.8$^{\rm s}$, 
34$^{\circ}$ 18$^{\prime}$ 38$^{\prime\prime}$) \citep{vanderhost2005a, vanderhost2005b}.
The location of this radio source does not match with two possible 
X-ray sources associated with GRB 050925 and discussed in section 2.2.

\subsection{Search for Persistent Hard X-ray Emission by INTEGRAL}

\begin{figure}[t]
\begin{center}
\includegraphics[height=16cm]{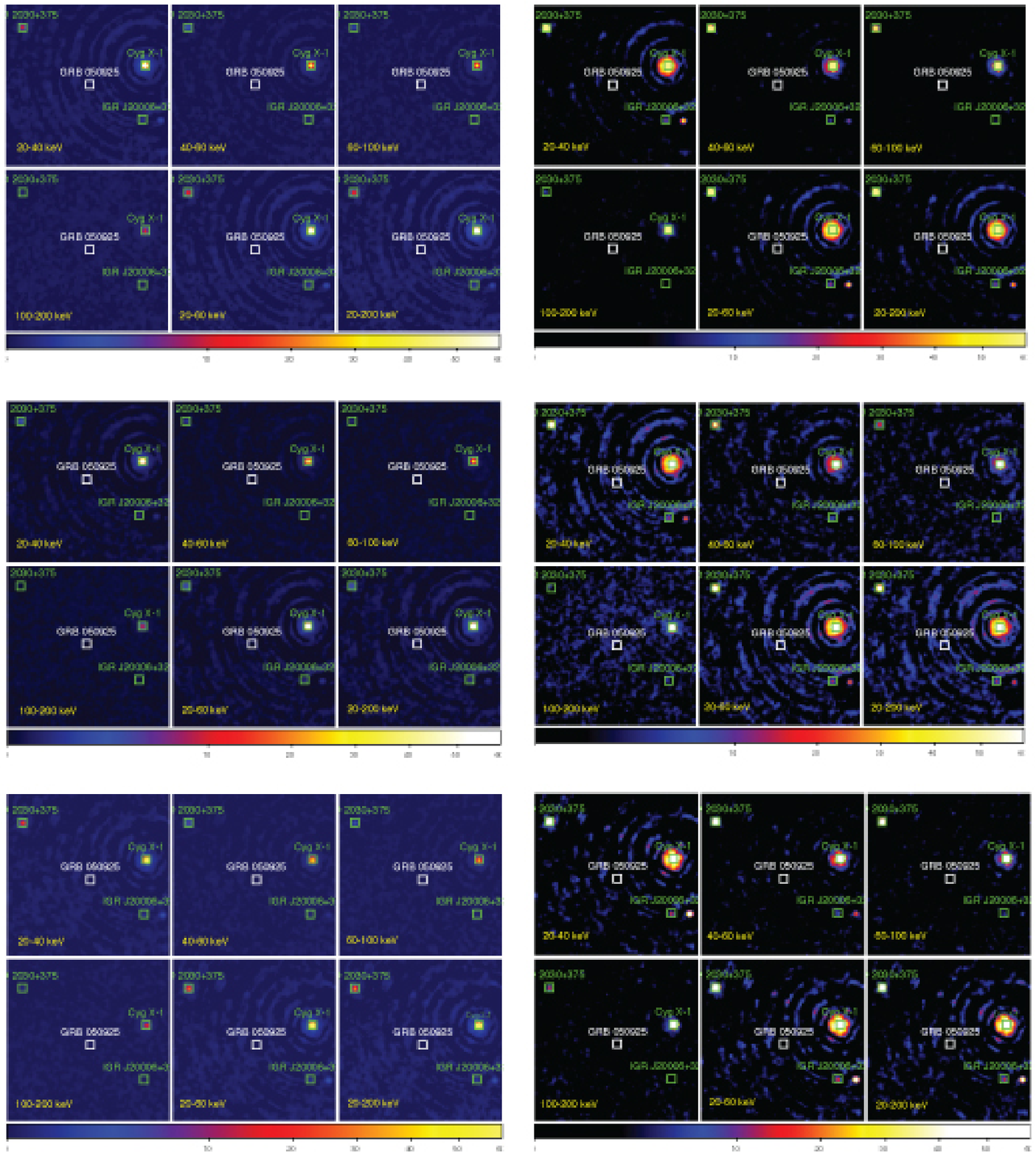}
\end{center}
\caption{The left column figures show the {\it INTEGRAL} ISGRI intensity maps of the total (top), 
the pre-burst (middle), and the post-burst (bottom) period.  The right column figures show 
the {\it INTEGRAL} ISGRI significance maps of the total (top), the pre-burst (middle) 
and the post-burst (bottom) period.  Each of the panels of the figure correspond to the 
different energy bands: the top panel from left to right: 20-40 keV, 40-60 keV and 60-100 keV band, 
and the bottom panel from left to right: 100-200 keV, 20-60 keV and 20-200 keV band.  
The location of GRB 050925 is shown in a white square.\label{isgri_maps}}
\end{figure}

The {\it INTEGRAL} data of the IBIS/ISGRI have been analyzed using the INTEGRAL 
Off-line Scientific Analysis package (OSA v7.0).  The data from December 2002 to May 2008, 
when the location of GRB 050925 is in the field of view of the IBIS, have been 
processed to cover the pre and the post burst periods of GRB 050925.  The ISGRI mosaic 
images are created for the pre-burst period (from December 21, 2002 to July 2, 2005), 
the post-burst period (from November 10, 2005 to May 23, 2008) and the total period 
(from December 21, 2002 to May 23, 2008) in the 20-40 keV, 40-60 keV, 60-100 keV, 
100-200 keV, 20-60 keV and 20-200 keV bands.  The total exposure of the mosaic 
images are 2.4 Ms, 2.5 Ms and 4.9 Ms for the pre-burst, the post-burst and the total 
period respectively.  The intensity maps and the significance 
maps of six energy bands and different periods are shown in 
Figure \ref{isgri_maps}.  The significance of the location of 
GRB 050925 is $<1.3$ $\sigma$ in all mosaic images.  Therefore, no persistent hard X-ray 
emission has found in the ISGRI data.  

Using the ISGRI sensitivity estimate of \citep{ibis_survey}, $f_{lim}^{5\sigma} = 0.77 
\times (T/Ms)^{-0.5}$ mCrab, we estimated the 5 $\sigma$ upper limits of the persistent 
hard X-ray emission at the location of GRB 050925 in the pre-burst, the post-burst and 
the total period as 0.50 mCrab, 0.49 mCrab and 0.35 mCrab respectively.  

\subsection{Search for short bursts by INTEGRAL}

To search for additional bursts from GRB 050925, we create 100 ms light curves in the 
20-40 keV, 40-60 keV, 60-100 keV, 100-200 keV, 20-60 keV and 20-200 keV bands using 
{\tt ii\_light} task of the OSA software package.  We set the criteria for a possible 
burst to have $>$7 $\sigma$ in a light curve bin in the 20-40 keV and the 40-60 keV bands 
simultaneously to avoid spurious short events which we usually only see in a 
single band\footnote{Since a short burst from a SGR has a spectrum from a few keV up 
to few 100 keV, we should see the event at light curves in a several continuous energy bands.}.  
We found no possible short burst at the location of GRB 050925 in our 4.9 Ms data.

\section{Discussion}
\begin{figure}
\begin{center}
\includegraphics[height=6cm,angle=-90,clip]{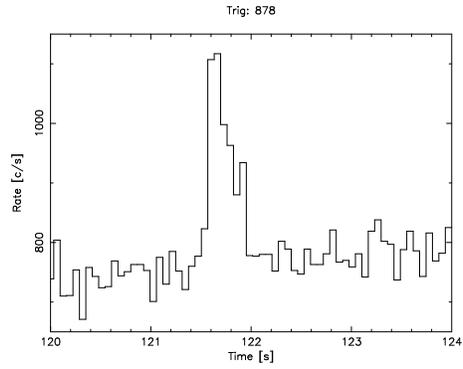}
\caption{\label{batse_trig878_lc} 64 ms light curve of the BATSE trigger 878.}
\end{center}
\end{figure}

We search the BATSE Gamma-Ray Burst Catalog\footnote{http://www.batse.msfc.nasa.gov/batse/grb/catalog/current/} 
for additional bursts from the location of GRB 050925.  We found seven events which their error radii 
match the location of GRB 050925.  Furthermore, we found three candidates (out of seven) 
which have the BATSE $T_{90}$ shorter than 2 s\footnote{We exclude one event, trigger ID 2468, 
because its duration information is not available in the catalog} (Table \ref{batse_candidates}).  
Figure \ref{batse_trig878_lc}-\ref{batse_trig1719_lc} 
shows the BATSE light curves of those events.  From the duration point of view, the BATSE trigger 
878 (4B 911007) has the shortest duration among the candidates.  However, its duration 
is still three times longer than that of GRB 050925.  Assuming these three triggers are associated 
with GRB 050925, the burst rate is estimated to be 0.33 yr$^{-1}$ (the BATSE catalog contains the 
events detected in 9 years of its operation).  Therefore, there is no surprise that we do not find 
any additional burst in the 4.9 Ms INTEGRAL data.  

Around the galactic longitude of $l = 70^{\circ}$, there is a complex of H$_{\rm II}$ regions (W 58) 
at a distance of 8.8 kpc \citep{georgelin}.  Assuming GRB 050925 is associated with the 
same H$_{\rm II}$ regions and therefore at a distance of 8.8 kpc , we discuss the source frame 
properties of the burst and its possible X-ray counterparts.  The total radiated energy is 
$7 \times 10^{38}$ ergs and the peak luminosity is $1 \times 10^{40}$ erg s$^{-1}$ for the burst.  
Therefore, the radiated energy and the peak luminosity are similar to those of weak short bursts from 
known SGRs.  Assuming the same distance, the X-ray luminosities of Src \#1 and Src \#4 during the XRT 
observation are $9 \times 10^{32}$ erg s$^{-1}$  and $3 \times 10^{32}$ erg s$^{-1}$ respectively.  
Since the X-ray afterglow luminosities of the SGR short bursts are 10$^{34}-10^{35}$ erg s$^{-1}$ 
\citep{nakagawa2008}, the X-ray luminosity of Src \#1 could be more consistent compared to 
that of Src \#4.  However, note that the X-ray luminosity of Src \#1 is still order of magnitude 
lower than the typical X-ray afterglow luminosities of the SGR short bursts.  
We also want to note that the shallow decay index ($<-1$) which we found for both Src \#1 and Src \#4 
is common for the X-ray afterglow of SGR short bursts \citep{nakagawa2008}.  

Assuming the distance of 8.8 kpc, the upper limit of the hard X-ray luminosity will be 
$<3 \times 10^{34}$ erg s$^{-1}$.  According to the detections of the hard X-ray emission 
of five magnetars by $INTEGRAL$ \citep{gotz2006}, the hard X-ray luminosities are in the 
range of $5 \times 10^{34}$ - $4 \times 10^{35}$ erg s$^{-1}$.  Therefore, our upper limit 
at the location of GRB 050925 might not deep enough to conclude 
a non-existence of the hard X-ray emission.  We probably need to revisit the $INTEGRAL$ data 
in the future with more data or to wait for the data from the future focusing hard X-ray mission 
$NuSTAR$ to answer this question.  

We would like to thank the anonymous reviewers for comments and suggestions that materially 
improved the paper.  

\begin{table}[t]
\caption{\label{batse_candidates}Three candidates found in the BATSE catalog.}
\centerline{
\begin{tabular}{ccccccc}\hline
GRB name   & ID   & Trigger time & R.A.  & Dec. & Error radius & $T_{90}$\\
           &      &    (UT)      & (deg) & (deg) & (deg)       & (s)    \\\hline
4B 911007  & 878  & 15:32:09.2 & 308.91 & 40.59 & 19.67 & 0.384\\
4B 920722  & 1719 & 21:21:27.2 & 321.19 & 36.75 & 15.71 & 1.047\\
4B 930219E & 2205 & 20:54:06.0 & 302.41 & 20.02 & 15.98 & 1.152\\\hline
\end{tabular}
}
\end{table}

\begin{figure}
\begin{center}
\includegraphics[height=6cm,angle=-90,clip]{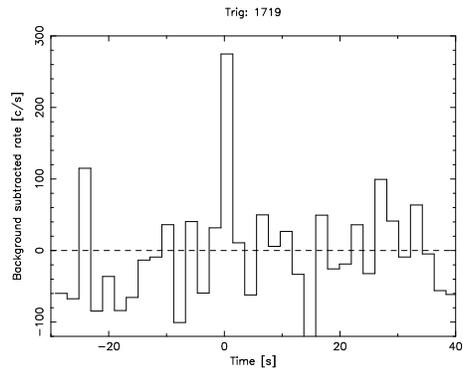}
\caption{\label{batse_trig1719_lc} 1 s light curve of the BATSE triger 1719.}
\end{center}
\end{figure}
\begin{figure}
\begin{center}
\includegraphics[height=6cm,angle=-90,clip]{batse_2205_64ms_tot_lc.ps}
\caption{\label{batse_trig1719_lc} 64 ms light curve of the BATSE triger 2205.}
\end{center}
\end{figure}








\end{document}